\documentclass[amssymb,prd,aps,showpacs,nofootinbib,twocolumn,]{revtex4}

\usepackage{latexsym}
\usepackage{bm}

\usepackage{amsmath}
\usepackage{graphicx}
\usepackage[T1]{fontenc}

\makeindex
\begin{document}

\title{The presence of Primordial Gravitational Waves in the Cosmic Microwave Background}

\author{Wytler Cordeiro dos Santos}
\email{wytler@fis.unb.br}
\affiliation{{\small Faculdade UnB Gama, Universidade de
Bras\'\i lia, CEP 72444-240, Gama, DF, Brasil}}


\begin{abstract}
\noindent The General Relativity affirms that any field is a source of gravitational field,
thus one should affirm that the energy of Cosmic Microwave Background (CMB) 
generated primordial gravitational waves. The present article shows that a gravitational wave with dimensionless amplitude $\sim 10^{-5}$ and large  wave length $\sim 10$ megaparsecs shifts temperature of CMB radiation about of a part in $10^{5}$.

\end{abstract}
\date{\today }

\maketitle

\section{Introduction}
The existence of the Cosmic Microwave Background (CMB) provides crucial support for the 
hot Big Bang cosmological model. The CMB visible today was once in thermal equilibrium
with the primordial plasma of the early universe, when the universe at that time was highly
uniform. However, the universe could not have been perfectly uniform at that time or no structures
would have formed subsequently. 
The careful investigation of small temperature and polarization fluctuations in the CMB, reflecting
small variation in density and velocity in the early universe \cite{Kosowsky1}.
The temperature fluctuations in the CMB are due to three effects: a change in the intrinsic
temperature of radiation at a given point in space, a Doppler shift if the radiation at a particular
point is moving with respect to the observer and a difference in gravitational potential
between a particular point in space and an observer \cite{Kamionkowski1,Gawiser,Sachs,White}
In addition to temperature fluctuations, the CMB is polarized. 
The polarization fluctuations of the CMB are expected to be significantly smaller than the temperature fluctuations and there are two types of polarization: 
electric modes (E-modes) produced by scalar perturbation from Thomson scattering \cite{Kosowsky1}
and magnetic modes (B-modes) that were not produced from the primordial plasma alone and  
they are signal from cosmic inflation and suggest the presence of gravitational waves \cite{Kamionkowski1,Gawiser,Zaldarriga}.
Thus, by analogy to CMB, there are of cosmological origin, a stochastic background of 
primordial gravitational waves (PGW) produced in the early universe. The PGW is regarded as
a strong evidence for the inflationary models, which is directly related to the energy scale
of inflation. They would carry informations on the state of the very early universe and on 
high-energies physics \cite{Maggiore,Allen,Andersson,Corda}.

It is acceptable the opinion based on firm evidences that CMB and PGW are directly related \cite{Grishchuk,Grishchuk2,Zhao1,Zhao2,Baskaran1,Baskaran2,Corda2}. Also, there are many possible processes during the very early stages of the universe, including quantum fluctuations during inflation, Inflaton field, bubble collisions in a first-order phase transition, the decays of cosmic strings and the processes that acted as seeds for galaxy formation would generate gravitational waves \cite{Maggiore,Allen,Andersson,Corda}.
Should be possible that in the very early universe the strong gravitational field 
interacting with electromagnetic field did produce PGW and
CMB at the same time? If we consider that the CMB
filling completely the space of our universe is a possible source of gravitational waves
that permeate the cosmos, we should affirm that CMB and PGW are tied.
The General Relativity states that any amount of energy is a source of gravity,
so the energy of CMB is the source of some  gravitational field, and it is asserted in Refs. \cite{Torre,Wils} that homogeneous pure radiation solutions to the Einstein equations admit electromagnetic sources. The subject of this work
is to demonstrate that the energy density of CMB, although it is very small, would generate contribution for stochastic background of gravitational waves of cosmological origin. 
 

We assume spacetime with Lorentzian metric tensor  $g_{\mu\nu}$, with  signature $(+---)$ \cite{Hobson,Landau}. Lower case
Greek indices refer to coordinates of spacetime and take the
values $0,1,2,3.$ The relation between the metric field
$g_{\mu\nu}$ and the material contents of spacetime is expressed
by Einstein's field equation,
\begin{equation}
\label{field equation}
G_{\mu\nu}=R_{\mu\nu}-\frac{1}{2}g_{\mu\nu}R=\kappa T_{\mu\nu}-\Lambda g_{\mu\nu},
\end{equation}
$T_{\mu\nu}$ being the stress-energy-momentum tensor, $\Lambda$ is cosmological constant, $R_{\mu\nu}$ the contract curvature tensor (Ricci tensor) and $R$ its trace.
This paper employs SI units with $\kappa= \dfrac{8\pi G}{c^4} $.


\section{Gravitational Waves}

Gravitational waves produced in the early universe would form a stochastic background which can span a very wide range of frequencies. The intensity of a stochastic gravitational wave background is characterized by the dimensionless quantity
\begin{equation}
\Omega(f)=\frac{1}{\rho_c}\frac{d\rho_{\mbox{\tiny GW}}}{d\log f}
\end{equation}
where $\rho_{\mbox{\tiny GW}}$ is the energy density of the gravitational waves, $f$ is the frequency and $\rho_c$ is the critical energy density for closing the universe. The critical density is given as a function of the present value of the Hubble constant $H_0$ such as $\rho_c=\frac{3c^2H_0^2}{8\pi G}$. A particular gravitational wave detector would measure the dimensionless gravitational wave strain
\begin{equation}
\label{h_c}
h_c\approx 1.3\times 10^{-20}\sqrt{\Omega(f)h_0^2}\left(\frac{100\mbox{ Hz}}{f}\right),
\end{equation}
where $h_0$ is the rescaled Hubble constant, expected to lie in the range $0.4\lesssim h_0\lesssim 0.9$. For example, for $\Omega(f) \sim 10^{-8}$ at a frequency of 100 Hz the strain in a gravitational wave detector is $h_c\sim 10^{-24}$ \cite{Andersson}.

CMB data shows the observable universe to be nearly flat \cite{Adler, Knox, Weeks, Aurich}. The flatness of the observable universe  requires that a stochastic background of PGW have not curved substantially the space. 
The first aim of this paper is an adequate choice for $g_{\mu\nu}$ that can represent and describe how PGW have connection with CMB with no relevant change for large scale structure of spacetime. In spite of this effort one can use exponential metric fields \cite{Santos},
\begin{equation}
\label{exp}
g_{\mu\nu}=e^{\Phi}\eta_{\mu\nu}+\sinh\Phi\Upsilon_{\mu\nu}
\end{equation}
where $\Phi$ is the strength of gravitation that can be weighed to match up to  $h_c$ from (\ref{h_c}) and therefore one could require that $\Phi$ is a constant amplitude of PGW.
The $\eta_{\mu\nu}=\mbox{diag}(1,-1,-1,-1)$ is the simplest empty spacetime:
the Minkowski spacetime, the flat universal covering space for all such derived spaces
\cite{Hawking}.  The conformal structure of Minkowski space is
what one would regard as the `normal' behavior of a spacetime at
infinity. 
The $\Upsilon_{\mu\nu}$
is a tensor on a background Minkowski spacetime, similar to
deviation $h_{\mu\nu}$ from linearized version of general
relativity. But instead one has the infinitesimal
condition for $|h_{\mu\nu}|\ll 1$, it is accepted that the
magnitude of $\Upsilon_{\mu\nu}$ can be equal to the magnitude of
empty flat spacetime ($|\Upsilon_{\mu\nu}|\approx |\eta_{\mu\nu}|
$). Moreover, it might be defined as an important mathematical
relationship among $\Upsilon_{\mu\nu}$ themselves,
\begin{eqnarray}
\label{ID}
\Upsilon_{\mu\nu}\Upsilon^{\nu\rho}=-2{\Upsilon_{\mu}}^{\rho}.
\end{eqnarray}
Thus, let us consider the simplest case of a monochromatic gravitational wave of one frequency $\omega=ck_G$ (where $k_G$ is the wave number) with amplitude $\Phi$ traveling in the positive $z$-direction. So, one can impose restrictions to $g_{\mu\nu}$, such as $g_{00}=1$ and $g_{33}=-1$, so that
\begin{equation}
\label{exp2}
g_{ab}=-e^{\Phi}\delta_{ab}+\sinh\Phi\Upsilon_{ab},
\end{equation}
with $a=1,2$ and $b=1,2$ whose tensor $\Upsilon_{ab}$ can be performed to produce circularly polarized gravitational wave,
\begin{equation}
(\Upsilon_{ab})=\begin{pmatrix}
1+\cos\theta_{G} & \sin\theta_G\cr
\sin\theta_G  & 1-\cos\theta_G
               \end{pmatrix}
\end{equation}
obeying the identity (\ref{ID}) to which
\begin{equation}
\label{U}
\Upsilon_{ab}\Upsilon^{bc}=-2{\Upsilon_a}^{c}
\end{equation}
is valid. The argument $\theta_G$ denotes the gravitational wave moving in the positive $z$-direction such as $\theta_G=k_G(z-ct)$.
Also, one can use
\begin{equation}
g^{ab}=-e^{-\Phi}\delta^{ab}-\sinh\Phi\Upsilon^{ab}
\end{equation}
for the evaluation the components of inverse of the matrix $(g_{ab})$.
Finally, the way that ends this reasoning is expressing the covariant components of metric tensor as
\begin{equation}
\label{metrica}
g_{\mu\nu}\Rightarrow
\begin{cases}
g_{00}=1\cr
g_{11}=-\cosh\Phi+\sinh\Phi\cos\theta_G\\
g_{12}=g_{21}=\sinh\Phi\sin\theta_G\\
g_{22}=-\cosh\Phi-\sinh\Phi\cos\theta_G\\
g_{33}=-1
\end{cases},
\end{equation}
and contravariant components of the metric tensor given by
\begin{equation}
\label{metrica_inversa}
g^{\mu\nu}\Rightarrow
\begin{cases}
g^{00}=1\cr
g^{11}=-\cosh\Phi-\sinh\Phi\cos\theta_G\\
g^{12}=g^{21}=-\sinh\Phi\sin\theta_G\\
g^{22}=-\cosh\Phi+\sinh\Phi\cos\theta_G\\
g^{33}=-1
\end{cases}.
\end{equation}
An interesting observation is the spacetime of this circularly polarized plane wave with volume element $\sqrt{-g}d^4x$ is the same of Minkowski spacetime. In this sense this gravitational wave does not modify flatness of background Minkowski spacetime where this plane wave travels onto.

For small deviations from
flat spacetime  where $\Phi\ll 1$, the components $g_{ab}$  of metric tensor (\ref{exp2}) should be denoted in the linear approximation as
\begin{equation}
g_{ab}\approx (1-\Phi)\delta_{ab} +\Phi\Upsilon_{ab}=\delta_{ab}+\Phi(\Upsilon_{ab}-\delta_{ab}),
\end{equation}
where the above equation allows rewrite $g_{\mu\nu}$ as
\begin{equation}
\label{linear}
g_{\mu\nu}=\eta_{\mu\nu}+h^{TT}_{\mu\nu},
\end{equation}
with
\begin{equation}
(h_{\mu\nu}^{TT})=\Phi\begin{pmatrix}
0&0&0&0\\
0&\cos[k_G(z-ct)]&\sin[(k_G(z-ct)]&0\\
0&\sin[(k_G(z-ct)]&-\cos[(k_G(z-ct)]&0\\
0&0&0&0
           \end{pmatrix}\nonumber
\end{equation}
corresponding a weak gravitational field of a circularly polarized plane gravitational wave propagating in the $z-$direction in the TT gauge.
The linearized gravitational theory ignores the energy-momentum associated with the gravitational wave itself \cite{Hobson}. To include this contribution, and thereby go beyond the linearized theory, one must modify field equations to read $G^{(1)}_{\mu\nu}$ from (\ref{linear}),
\begin{equation}
G^{(1)}_{\mu\nu}=\kappa \left(T_{\mu\nu}+T^{(GW)}_{\mu\nu}\right),
\end{equation}
where $G^{(1)}_{\mu\nu}$ is the linearized Einstein tensor, $T_{\mu\nu}$ is the energy-momentum tensor of any matter or any field present and $T^{(GW)}_{\mu\nu}$ is the energy-momentum tensor of gravitational field itself. However, one may expand beyond first order to obtain
\begin{equation}
G_{\mu\nu}=G^{(1)}_{\mu\nu}+G^{(2)}_{\mu\nu}+\cdots = \kappa T_{\mu\nu},
\end{equation}
where superscript in parentheses indicate the order of the expansion in $h_{\mu\nu}$, to suggest that with good approximation, one should make the identification
\begin{equation}
T^{(GW)}_{\mu\nu} = -\kappa^{-1}G^{(2)}_{\mu\nu}.
\end{equation}
Thus, in this linearized theory, the energy-momentum tensor can be written as
\begin{equation}
\label{TGW}
T^{(GW)}_{\mu\nu}=\frac{c^2}{16\pi G}\omega_{G}^2\Phi^2 L_{\mu\nu},
\end{equation}
where
\begin{equation}
\label{L}
(L_{\mu\nu})=
 \begin{pmatrix}
1&0&0&-1\cr
0&0&0&0\cr
0&0&0&0\cr
-1&0&0&1     \end{pmatrix},
\end{equation}
with the flux in the direction of propagation
\begin{equation}
F=cT_{00}^{(GW)}=\frac{c^3}{16\pi G}\omega_{G}^2\Phi^2.
\end{equation}


It is important to give special emphasis on the exponential metric fields (\ref{exp2}). Because of relationship among $\Upsilon_{ab}$ themselves (\ref{U}), this tensor has structure limited in the first order. 
So, the non-linearity of $g_{\mu\nu}$ is less hard, at least in a tensorial description way.
And differently from linearized version of gravity 
we must use (\ref{metrica}) in order to obtain the effective components $R_{\mu\nu}$ of the Ricci tensor for any order
\begin{equation}
\label{R}
R_{\mu\nu}=\frac{(k_G)^2}{2}\sinh^2\Phi \,\,L_{\mu\nu}.
\end{equation}
The non-zero components of the Ricci tensor ${R^{\mu}}_{\nu}$ are 
${R^{0}}_{0}=-{R^{3}}_{3}$, therefore one finds that the curvature scalar results in
$R=g^{\mu\nu}R_{\mu\nu}={R^{\mu}}_{\mu}=0$. So the components of Einstein tensor are 
\begin{equation}
\label{Gmunu}
G_{\mu\nu}=\frac{\omega_G^2}{2c^2}\sinh^2\Phi \,\,L_{\mu\nu},
\end{equation}
that satisfy the classification of homogeneous pure radiation solutions given in Ref. \cite{Stephani}.

One can expand $G_{\mu\nu}$ in powers of $\Phi$,
\begin{eqnarray}
\label{G_expand}
G_{\mu\nu}&=&\frac{\omega_{G}^2}{2c^2}\left(\Phi^2+\frac{\Phi^4}{3}+\frac{2\Phi^6}{45}+\cdots\right)L_{\mu\nu}\cr
&=&G^{(2)}_{\mu\nu}+G^{(4)}_{\mu\nu}+G^{(6)}_{\mu\nu}+\cdots ,
\end{eqnarray}
to observe that $G^{(1)}_{\mu\nu}=0$ (solution of the linearized field equations in vacuum) and the others odd-orders are nulls.
The second-order of $G_{\mu\nu}$ is 
\begin{equation}
G^{(2)}_{\mu\nu}=\frac{\omega^2_{G}}{2c^2}\Phi^2L_{\mu\nu},
\end{equation}
that results in $T_{\mu\nu}^{(GW)}$ of the identification (\ref{TGW}), one of the terms that should compose energy-momentum tensor of gravitational wave.

So, the effective result (\ref{Gmunu}) obtained from metric field (\ref{metrica}) should be connected with material or radiation contents of spacetime.

\section{The connection between Electromagnetic Radiation and Primordial Gravitational Waves}

In a general cosmology model the universe is assumed to contain matter, radiation and the vacuum. One usually adopts the viewpoint that the cosmological fluid consist of three components each with a different equation of state. 
In this sense, one usually can affirm that the large scale background is 
\begin{equation}
\label{field_equations}
G_{\mu\nu}=\kappa\left(T^{(\mbox{matter})}_{\mu\nu}+T^{(\mbox{radiation})}_{\mu\nu}\right)-\Lambda g_{\mu\nu}.
\end{equation}
For an amount empty space $T^{(\mbox{matter})}_{\mu\nu}=0$. One can also observe that cosmological constant, 
$\Lambda\sim 10^{-53}m^{-2}$, must be fine-tuned to match the density of universe \cite{Hobson}, such as for an amount of the empty space is acceptable to set $\Lambda\approx 0$.
There are various fields on space-time, such as the electromagnet field, neutrino field, etc., 
including gravitational waves which describes the radiation energy density of the universe.
It is worth noting that, to a very good approximation, the dominant contribution to the radiation energy density of the universe is due to the photons of CMB which makes up a fraction of roughly $\Omega_{rad,0}\approx 5\times 10^{-5}$ of the total density of the universe \cite{Hobson,Liddle}. In this case, we find that
\begin{equation}
T^{(\mbox{radiation})}_{\mu\nu}=T^{(ED)}_{\mu\nu},
\end{equation}
where $T^{(ED)}_{\mu\nu} $ is electrodynamic stress-energy-momentum for CMB.


Let us now consider that the CMB radiation is a source of gravitational waves
and, as in special relativity, one can introduce a vector potential $A^{\alpha}$ of an electromagnetic field tensor
\begin{equation}
\label{Fuv}
F_{\alpha\beta}=\nabla_{\alpha}A_{\beta} - \nabla_{\beta}A_{\alpha}=\partial_{\alpha}A_{\beta} - \partial_{\beta}A_{\alpha},
\end{equation}
and one can derive the wave equation that governs the vector potential in the vacuum, obtaining Maxwell's equation at vacuum
\begin{equation}
\label{ED_eq1}
\nabla_{\beta}F^{\alpha\beta}=\nabla_{\beta}\nabla^{\alpha}A^{\beta}-\nabla_{\beta}\nabla^{\beta}A^{\alpha}+{R^{\alpha}}_{\beta}A^{\beta}=0.
\end{equation}
Adopting the standard approach of special relativity, impose the Lorentz gauge condiction
\begin{equation}
\nabla_{\beta}A^{\beta}=\frac{1}{\sqrt{-g}}\partial_{\beta}\left(A^{\beta}\sqrt{-g}\right)=0
\end{equation}
that in this spacetime (\ref{metrica}),  $\sqrt{-g}=1$ and one can then write
\begin{equation}
\label{gauge_cond}
\partial_{\beta}A^{\beta}=0,
\end{equation} 
thereby bringing the wave equation (\ref{ED_eq1}) into the form
\begin{equation}
\label{ED_eq2}
-\nabla_{\beta}\nabla^{\beta}A^{\alpha}+{R^{\alpha}}_{\beta}A^{\beta}=0.
\end{equation}
In situations where the spacetime scale of variation of electromagnetic field is much 
smaller than that of the curvature, one would expect to have solutions of Maxwell's equations of the form 
of a wave oscillating with nearly constant amplitude, i.e., solutions of the form
\begin{equation}
\label{bicho_1}
A^{\alpha}=C^{\alpha}e^{i\xi},
\end{equation}
where derivatives of $C^{\alpha}$ are small. Substituting above equation into equation (\ref{ED_eq2}) and neglecting the small term $\nabla_{\beta}\nabla^{\beta}C^{\alpha}$ as well as the Ricci tensor term yields the condiction
\begin{equation}
\nabla_{\beta}\xi\nabla^{\beta}\xi=0,
\end{equation}
 i.e., we find that the surfaces of constant phase are null, and thus 
(by the same argument as given for flat space time) $k_{\alpha}=\nabla_{\alpha}\xi$ 
is tangent to null geodesics. This suggest that, in this `geometrical optics approximation', light travels on null geodesics \cite{Wald}.

After one has obtained a solution as (\ref{bicho_1}), it is possible to calculate 
electromagnetic field tensor with potential vector $A^{\alpha}$ for the CMB radiation.
The simplest electromagnetic field is that of a plane wave; so,
the gauge condiction (\ref{gauge_cond}) indicates that the potential vector propagating on $z$-direction is given by
\begin{equation}
\label{luz}
A^{\alpha} =\begin{pmatrix}
0\\
a\sin(\theta_E+\delta_1)\\
b\sin(\theta_E+\delta_2)\\
0
                          \end{pmatrix},
\end{equation}
where $\theta_E$ denotes the variable part of the phase factor, i.e. $\theta_E=k_E(z-ct)$, such as $k_E$ is the wave-length of electromagnetic wave, $a$ and $b$ are the maximum
amplitudes and $\delta_1$  and $\delta_2$ are arbitrary phases. 
In general, the above optical field is elliptically polarized, but there are several combinations of amplitude and phase: (i) if $a=0$ (or $b=0$) we have linearly horizontal (or vertical) polarized light; (ii) if $a=b$ and $\delta_2-\delta_1=0$
(or $\delta_2-\delta_1=\pi$) we have linear polarized light; (iii) if $a=b$ and $\delta_2-\delta_1=\pm\pi/2$, we have right/left circularly polarized light.
So, it is necessary to calculate $F_{\alpha\beta}$ from expression (\ref{Fuv}), therefore that
$A_\alpha=g_{\alpha\beta}A^{\beta}$ with $A_0=0$, $A_1=g_{11}A^1+g_{12}A^2$, $A_2=g_{21}A^1+g_{22}A^2$ and $A_3=0$. In fact, the term ${R^{\alpha}}_{\beta}A^{\beta}$ in equation (\ref{ED_eq2}) is vanished by (\ref{R}) and (\ref{L}).

Thus, the only components non-vanish of $F_{\alpha\beta}$ are, 
\begin{eqnarray}
F_{01}=F_{13}=-F_{10}=-F_{31}\cr
F_{02}=F_{23}=-F_{20}=-F_{32},
\end{eqnarray}
with
\begin{small}
\begin{eqnarray}
\label{F01}
F_{01}&=&\partial_0A_1-\partial_1A_0=\frac{1}{c}\frac{\partial}{\partial t}\left(g_{11}A^1+g_{12}A^2\right)\cr
F_{01}&=& ak_E\cos(\theta_E+\delta_1)\sinh\Phi\times \cr
& &\left[\left(\frac{k_E}{k_G}\right)\sin\theta_G\tan(\theta_E+\delta_1)-\cos\theta_G+\coth\Phi\right]\cr
&-& bk_E\cos(\theta_E+\delta_2)\sinh\Phi\times\cr 
& &\left[\left(\frac{k_G}{k_E}\right)\cos\theta_G\tan(\theta_E+\delta_2)+\sin\theta_G\right].
\end{eqnarray}
\end{small}
The electromagnetic radiation of CMB has a Planck blackbody spectrum with caracteristic frequency 
$f_0=5.7\times 10^{10} Hz$.
One the most exciting prospect in the study of the stochastic background of gravitational radiation is the possibility of detecting gravitational radiation produced at the very early epoch of the universe, $t\approx 10^{-22}$ sec. At that early epoch with an energy of order 10 MeV, the LIGO experiment would detect gravitational radiation with frequency of order 100 Hz.
Gravitational radiation produced at time of order $t\approx 10^{-14}$ sec after the Big Bang would have at the present day frequency $\approx 10^{-2}$ Hz, that may be detected by LISA experiment \cite{Allen}.
So, we do expect that primordial gravitational waves have values
$k_G\ll k_E\Rightarrow \theta_G\ll\theta_E$, such that
\begin{eqnarray}
\label{id_trig}
\cos(\theta_E+\delta_i)\cos\theta_G &\approx& \cos(\theta_E+\delta_i) \hspace*{1cm}\mbox{and}\cr
\cos(\theta_E+\delta_i)\sin\theta_G &\approx& 0,
\end{eqnarray}
where, one finally gets to $F_{01}$ from (\ref{F01}),
\begin{equation}
F_{01}\approx ak_E \cos(\theta_E+\delta_1)e^{-\Phi}.
\end{equation}
Also, one can calculate $F_{02}$ and gets
\begin{equation}
F_{02}\approx bk_E\cos(\theta_E+\delta_2)e^{\Phi}.
\end{equation}
So, the result to electromagnetic field-strength tensor  $F_{\alpha\beta}$ is
\begin{equation}
(F_{\alpha\beta})=
	\begin{pmatrix}
0&F_{01}&F_{02}&0\\
-F_{01}&0&0&F_{01}\\
F_{02}&0&0&F_{02}\\
0&-F_{01}&-F_{02}&0
             \end{pmatrix}
\end{equation}
The stress-energy-momentum tensor of electromagnetic field is given by (in SI units \cite{Hobson}), 
\begin{equation}
\label{EM_tensor}
T_{\mu\nu}^{(ED)}=\frac{1}{\mu_0}g^{\alpha\beta}F_{\mu\alpha}F_{\beta\nu}-\frac{1}{4\mu_0}g_{\mu\nu}F_{\alpha\beta}F^{\alpha\beta},
\end{equation}
where the constant $\mu_0$ is the permeability of free space. 
This stress-energy-momentum tensor
acts as the source of gravity in Einstein's field equation. It therefore plays an important part in any attempt to model the dynamics involving the gravitational field coupled to electromagnetic field.

One can calculate $g^{\alpha\beta}F_{\beta\nu}={F^{\alpha}}_{\nu}$ and finds that
\begin{small}
\begin{equation}
({F^{\alpha}}_{\nu})=
	\begin{pmatrix}
0&F_{01}&F_{02}&0\\
-g^{11}F_{01}-g^{12}F_{02}&0&0&g^{11}F_{01}+g^{12}F_{02}\\
-g^{12}F_{01}-g^{22}F_{02}&0&0&g^{12}F_{01}+g^{22}F_{02}\\
0&F_{01}&F_{02}&0
                   \end{pmatrix},
\end{equation}
\end{small}
and so the first term of $T_{\mu\nu}^{(ED)}$ becomes,
\begin{equation}
F_{\mu\alpha}{F^{\alpha}}_{\nu}=-\left[g^{11}(F_{01})^2-2g^{12}F_{01}F_{02}+g^{22}(F_{02})^2\right]L_{\mu\nu}.
\end{equation}

With contravariant components of the metric tensor, $g^{\mu\nu}$ of (\ref{metrica_inversa}) and the results of (\ref{id_trig}) one can find,
\begin{small}
\begin{equation}
F_{\mu\alpha}{F^{\alpha}}_{\nu}\approx k_E^2[a^2\cos^2(\theta_E+\delta_1)e^{-\Phi}
+b^2\cos^2(\theta_E+\delta_2)e^{\Phi}]L_{\mu\nu}.
\end{equation}
\end{small}
Observe that the second term of $T_{\mu\nu}^{(ED)}$ from (\ref{EM_tensor}) vanishes, because it is proportional to $F_{\alpha\beta}F^{\alpha\beta}$ that is proportional to $L_{\mu\nu}g^{\mu\nu}=0$. 
So the stress-energy-momentum tensor of electromagnetic field is
\begin{small}
\begin{equation}
T_{\mu\nu}^{(ED)}=\frac{1}{\mu_0}k_E^2[a^2\cos^2(\theta_E+\delta_1)e^{-\Phi}
+b^2\cos^2(\theta_E+\delta_2)e^{\Phi}]L_{\mu\nu}.
\end{equation}
\end{small}
Thus we can use the fact that averaged over several wavelengths, 
\begin{equation}
\langle \cos^2(\theta_E+\delta_i) \rangle=\frac{1}{2},
\end{equation}
the energy-momentum tensor reads,
\begin{equation}
\langle T_{\mu\nu}^{(ED)}\rangle=\frac{k_E^2}{2\mu_0}[a^2e^{-\Phi}
+b^2e^{\Phi}]L_{\mu\nu}.
\end{equation}
Now, we may define:
\begin{eqnarray}
\label{epson}
a&=&{\cal A} \cos\epsilon \hspace*{1cm}\mbox{and}\cr
b&=&{\cal A} \sin\epsilon,
\end{eqnarray}
thus, the energy-momentum tensor is,
\begin{equation}
\label{ED_tensor}
\langle T_{\mu\nu}^{(ED)}\rangle=\frac{k_E^2{\cal A}^2}{2\mu_0}\left[\cosh\Phi-\cos(2\epsilon)\sinh\Phi\right]L_{\mu\nu}.
\end{equation}
If there are no gravitational field then one can set $\Phi=0$ and we have the electromagnetic energy-momentum tensor in Minkowski space time,
\begin{equation}
\langle T_{\mu\nu}^{(ED)}\rangle=\frac{k_E^2{\cal A}^2}{2\mu_0}L_{\mu\nu},
\end{equation}
with energy flux $I$ (W/m$^2$) in the $z$-direction given by,
\begin{equation}
\label{flux}
I=c\langle T_{00}^{(ED)}\rangle,
\end{equation}
that we may choose the energy flux for CMB, given by Stefan-Boltzmann law 
\begin{equation}
\label{flux2}
I^{\mbox{\tiny(CMB)}}_0=\sigma T^4=3.15\times 10^{-6} \mbox{W/m}^2,
\end{equation}
where $\sigma$ is the Stefan-Boltzmann constant and $T$ is the temperature ($T\approx 2.73$ K) of the photons of CMB.
Thus, we can denote electromagnetic energy-momentum tensor (\ref{ED_tensor}) by,
\begin{equation}
\label{ED_tensor2}
\langle T_{\mu\nu}^{(ED)}\rangle=\frac{I^{\mbox{\tiny(CMB)}}_0}{c}\left[\cosh\Phi-\cos(2\epsilon)\sinh\Phi\right]L_{\mu\nu}.
\end{equation}
One can expand $\langle T_{\mu\nu}^{(ED)}\rangle $ to the first order of $\Phi$,
\begin{equation}
\label{ED_tensor3}
\langle T_{\mu\nu}^{(ED)}\rangle\approx \frac{I^{\mbox{\tiny(CMB)}}_0}{c}\left[1-\cos(2\epsilon)\Phi\right]L_{\mu\nu}.
\end{equation}
Observe that intensity of CMB radiation in an arbitrary $z$ axis should float in function of polarized parameter $\epsilon$ of light field (\ref{luz}) and (\ref{epson}), and in function of amplitude of gravitational field $\Phi$.

It is straightforward to rewrite the gravitational field equations (\ref{field_equations}), by noting that
\begin{equation}
G_{\mu\nu}=\kappa\langle T^{(ED)}_{\mu\nu}\rangle,
\end{equation}
if one substitute the Einstein tensor from expression (\ref{Gmunu}) and electromagnetic energy-momentum tensor from (\ref{ED_tensor2}), we have the below relation that states that an amount of electromagnetic energy is the source of gravitational wave,
\begin{equation}
\frac{\omega_G^2}{2c^2}\sinh^2\Phi \,\,L_{\mu\nu}=\kappa\frac{I^{\mbox{\tiny(CMB)}}_0}{c}\left[\cosh\Phi-\cos(2\epsilon)\sinh\Phi\right]L_{\mu\nu},
\end{equation}
and assuming that $\Phi\ll 1$, the above equation results in
\begin{equation}
\Phi^2=\frac{8I^{\mbox{\tiny(CMB)}}_0}{h}\left(\frac{\ell_P}{f_G}\right)^2,
\end{equation}
where $h=6.63\times 10^{-34}$ J$\cdot$ s is the Planck' constant, $\ell_P=\sqrt{G\hslash/c^3}=1.62\times 10^{-35}$ m is Planck length and $f_G=\frac{\omega_G}{2\pi}$ is the frequency of gravitational wave in present-day at temperature $T=2.73K$ in equilibrium with CMB radiation. Hence one obtains
\begin{equation}
\label{Phi}
\Phi=\sqrt\frac{8I^{\mbox{\tiny(CMB)}}_0}{h}\left(\frac{\ell_P}{f_G}\right)=\frac{3.16\times 10^{-21}\,\,{\mbox{s}}^{-1}}{f_G}.
\end{equation}
The spectrum for amplitude $\Phi$ versus frequency is plotted in below figure.
\begin{center}
\begin{figure}[ht]
\begin{centering}
\rotatebox{-90}{\resizebox{5.8cm}{!}{\includegraphics{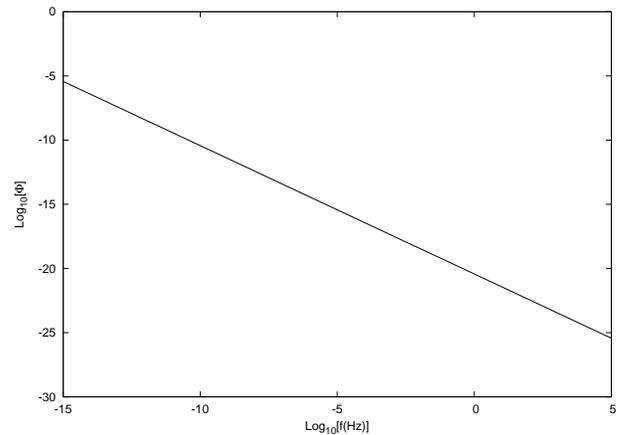}}}
\caption{Plot of amplitude $\Phi$ of gravitational wave versus
frequency $f$ given by expression (\ref{Phi}), valid to present-day with temperature $T=2.73K$. }
\label{none}
\end{centering}
\end{figure}
\end{center}
This spectrum seems reasonable similar to the graphics presented by \cite{Grishchuk,Grishchuk2, Baskaran1,Baskaran2}, for the frequency range from $10^{-15}$ Hz to $10^{5}$ Hz.


Note that this approach has allowed us to recognize that the expression (\ref{ED_tensor3}) leads us to suggest a flux fluctuation of the CMB radiation away from its value $I^{\mbox{\tiny(CMB)}}_0$. From equation (\ref{flux}) we have
\begin{equation}
\label{flux3}
I^{\mbox{\tiny(CMB)}}=I^{\mbox{\tiny(CMB)}}_0[1-\cos(2\epsilon)\Phi].
\end{equation}
In fact, as we know, the deviation from isotropy in the CMB radiation are of about one part in $10^5$, i.e. differences in the temperature of the blackbody spectrum measured in different direction in the sky, are of the order of $\delta T=20\mu K$ on large scales \cite{Allen,Garcia}. 
A temperature fluctuation $\delta T$ of the CMB radiation in (\ref{flux2}) results in
\begin{equation}
I^{\mbox{\tiny(CMB)}}=\sigma(T\pm \delta T)^4 \approx \sigma T^4\left(1\pm \frac{4\delta T}{T}\right),
\end{equation}
where it was retained terms only up to first order in $\delta T$, 
that reduces to 
\begin{equation}
\label{flux4}
I^{\mbox{\tiny(CMB)}}\approx I^{\mbox{\tiny(CMB)}}_0 (1\pm 10^{-5}).
\end{equation}
Now, we must compare the above expression with (\ref{flux3}). Observe that $|\cos(2\epsilon)|\leq 1$. If we have an amplitude of gravitational wave $\Phi\sim 10^{-5}$, the frequency of this wave is $f_G\sim 10^{-15}$ Hz, and one has a flux fluctuation of the CMB radiation to order of $10^{-5}$ given by (\ref{flux4}). This wave has a very large wavelength, $\lambda \approx  10$ Mpc. Thus, the effect of a long wavelength gravitational wave is to shift the temperature distribution of the CMB radiation on the celestial sphere away from perfect isotropy.


\section{Conclusion}

 There are indications of the presence of primordial gravitational waves in the CMB anisotropies \cite{Grishchuk,Corda2} and an easy way to approach it, may be one deals the energy density of CMB radiation, though it is very small, as a possible source for stochastic background of gravitational waves of cosmological origin. The spectrum obtained (\ref{Phi}) in this work, seems acceptable compared with (\ref{h_c}) from \cite{Andersson}, for example. The expression (\ref{h_c}) leads us to the amplitude gravitational wave of value $\sim 10^{-24}$ at a frequency of 100 Hz. In the meantime, the expression (\ref{Phi}) allows us to calculate an amplitude gravitational wave of value $\sim 10^{-23}$ at the same frequency of 100 Hz, that is acceptable. 
The connection between the temperature fluctuations of the CMB radiation and gravitational waves arises through the Sachs Wolfe effect \cite{Sachs}. Sachs and Wolfe showed how variations in the density of the cosmological fluid and gravitational wave perturbations result in CMB radiation temperature fluctuations. In accordance with them, this work shows that a gravitational wave of large wavelength (10 Mpc) shifts the temperature of CMB radiation about of a part in $10^5$.

\begin{acknowledgments}

The author would like to thank the REUNI (Reestrutura\c{c}\~ao e Expans\~ao das Universidades Federais), an institution whose founder is Luis In\'acio Lula da Silva.
\end{acknowledgments}
%
%
%

\end{document}